# Intra-operative quantification of the surgical gesture in orbital surgery: application to the proptosis reduction

V. Luboz[1], P. Swider[2], D. Ambard[2], F. Boutault[3,2], Y. Payan[1]

1. TIMC-IMAG Laboratory, UMR CNRS 5525, Faculté de Médecine, 38706 La Tronche, France
2. Biomechanics Laboratory EA 3697 / IFR30, Purpan University Hospital, 31059 Toulouse, France
3. Maxillofacial Surgery Department, Purpan University Hospital, 31059 Toulouse, France

Corresponding author:
Professor Pascal Swider
Biomechanics Laboratory EA 3697/IFR30
Purpan University Hospital
31059 Toulouse cedex
France
Tel: +33 (0)561 497 944
Fax: +33 (0)561 496 745
e-mail: swider@cict.fr

**Abstract:**

*Background.* Proptosis is characterized by a protrusion of the eyeball due to an increase of the orbital tissue volume. To recover a normal eyeball positioning, the most frequent surgical technique consists in the osteotomy of orbital walls combined with a loading on the eyeball to initiate tissue decompression. The first biomechanical models dealing with proptosis reduction, validated in one patient, have been previously proposed by the authors.

*Methods.* This paper proposed an experimental method to quantify the intra-operative clinical gesture in proptosis reduction, and the pilot study concerned one clinical case. The eyeball's backward displacement was measured by an optical 3D localizer and the load applied by the surgeon was simultaneously measured by a custom-made force gauge. Quasi-static stiffness of the intra-orbital content was evaluated.

*Findings.* The average values for the whole experiment was 16 N (SD: 3 N) for the force exerted by the surgeon and 9 mm (SD: 4 mm) for the eyeball backward displacement. The averaged quasi-static stiffness of the orbital content was evaluated to 2.4 N/mm (SD: 1.2) and showed a global decrease of 45% post-operatively.

*Interpretation.* The protocol and the associated custom-designed devices allowed loads, induced displacements and macroscopic stiffness of the orbital content to be measured intra-operatively. The clinical relevance has been demonstrated in a pilot study. To our knowledge, no study has been published allowing the clinical gesture in proptosis reduction to be quantified intra-operatively.

Associating an enlarged database and validated patient-related predictive models will reinforce the surgical efficiency and patient comfort contributing to diagnosis and intra-operative guidance.

*Keywords:* Orbital surgery, Proptosis, experimental method, computer assisted planning,

# 1. Introduction

Computer Assisted Surgery (CAS) aims at assisting surgeons to improve diagnosis, therapeutic gestures and follow-up implementing rational and quantitative approaches (Taylor et al., 1996). The ultimate goal is to increase safety and accuracy leading to a minimally invasive surgery for patient comfort. In this strategy, numerical models can assist surgical planning and techniques provided they are validated (Payan, 2005). Implementing patient-related models seemed to be a relevant strategy but a lack of knowledge in physiological boundary conditions and loads, and tissue material properties still drastically limit the reliability of predictive models (Viceconti et al., 2005).

Recent years showed the increase of studies dealing with the *in vitro* or *in vivo* tissue biomechanical characterization (Miller et al., 2000; Ottensmeyer et al., 2004; Gerard et al., 2005; Hendriks et al., 2006). Very recent studies concerned intra-operative measurement procedures (Jenkinson et al., 2005; Gosling et al., 2006), and this endeavoured to answer the need to better identify and quantify the clinical gesture. The clinical relevance of our approach has fallen in this framework relating to maxillofacial pathologies and more particularly proptosis reduction.

Proptosis is characterized by the increase of the volume of the orbital content mostly due to an endocrinal dysfunction (Saraux et al., 1987). The protrusion of the eyeball induces aesthetical problems and physiological disorders such as abnormal cornea exposition and pathological loading of the optic nerve, orbital blood vessels and ocular muscles. It may induce the alteration of visual acuity up to blindness. Once the endocrinal situation is stabilized, a surgical reduction of the proptosis is usually needed to decompress the orbital content. The commonly used technique is the "Bone Removal Orbital Decompression (BROD)" (Adenis et al., 2003). It aims at increasing the volume of the orbital cavity with an osteotomy (*i.e.* a bone resection) of the orbital walls (Stanley et al., 1989). The soft tissues are partially evacuated through the osteotomy to form a hernia and the surgeon may manually apply a controlled load on the eyeball to favour the decompression process (see Figure 1c). Limited cuts in the outer membrane containing the orbital soft tissues allow the physiological liquid to flow towards the maxillary and ethmoid sinus regions. The backward displacement of the eyeball to recover the normal position is initiated and it is fully reached after the complete reduction of orbital tissue inflammation.

This intervention is technically difficult because of the proximity of ocular muscles and the optic nerve, the tightness of the eyelid incision, and the narrowness of the operating field. The surgery must be minimally invasive and having specific tools to improve surgical planning could be very helpful. The authors proposed predictive models leading to clinical rules to assist in the pre-operative planning (Luboz et al., 2004, 2005). Correlations established between predicted results and pre- and post-operative clinical imaging allowed the reliability of the approach to be quantified in terms of geometry, volume, and kinematics.

The missing dimension was the dynamic aspect involving interrelated force and kinematics measures and such was the framework of the presented study where the goal was to investigate the clinical gesture intra-operatively. We hypothesized that measuring simultaneously, the load applied by the surgeon and the resulting kinematics of the eyeball could provide patient-related quantitative data to enrich the clinical knowledge in proptosis reduction. To proceed, a specific custom-made force gage coupled to a navigation station has been designed and evaluated in a clinical pilot study.

## 2. Material and Methods

The real-time process of data acquisition concerned the monitoring of the load exerted on the eyeball and the tracking of the induced displacement. A patented device and procedure derived from the study (Payan et al., 2005). The total mass of the device was 0.047 kg and the design, material choice and assembly guaranteed the sterilization procedure related to the medical instruments. The intra-operative data monitoring was achieved using a conventional laptop (250MHz processor, 64Mo Ram).

*2.1 Description of the measurement devices*

*2.1.1 Force measurement*

The force gage is shown in Figure 1. For patient security, a titanium elliptic corneal protection for laser treatment was intercalated between the device and the eyeball. The core of the device was the hollow aluminum cone (1) supporting the stainless steel membrane (2) clamped with the rigid ring (3). As shown in Figure 1a, the load was initially applied on the thin shell (4) and it was transmitted to the membrane (2) through the silicon cylinder (5). The strain into the membrane was measured by extensometric strain gages and converted into the resulting force $F_{z1}$ using a prior calibration procedure. The accuracy of 2% in the range [0 N-40 N] up to 40 Hz was established using a reference tensile machine (Elf 3200 Böse-Enduratec ®). The force sensor was monitored using a custom-made Labview program (National Instruments ®).

*2.1.2 Displacement measurement*

The eyeball displacement was measured using the Polaris system (NDI® Radolfzell, Germany) which is a 3D optical localizer device with an accuracy of 0.3 mm in the operating field. Thanks to infra red beam reflections monitored by two cameras, the system tracked the real-time relative positioning of the rigid body (6) shown in Figure 1 relative to the reference frame ($O_0$, $x_0$, $y_0$, $z_0$) of the surgical table in the operating room. The 3 displacements and the 3 rotations of the force gage were measured, and a matrix transform allowed the component $w_1$ of the force gage translation vector to be expressed in the local reference frame ($O_1$, $r_1$, $\theta_1$, $z_1$) (see Figure 1). The measured force $F_{z1}$ and the displacement $w_1$ were collinear and no motion of the patient head was verified during the procedure. The 3D localization was monitored using a custom-made Visual C++ ® program.

*2.1.3 Stiffness of the orbital content*

The loading velocity and acceleration imposed by the surgeon were low in time and limited in magnitude, and the mass of biological components was very weak so the kinetic energy and structural dissipated energy very slightly contributed to the energy balance of the orbital content. If the orbital complex could be considered as a one degree of freedom system at the macroscopic level, equation (1a) expressed the overall discrete stiffness $k_{z1}$ of the orbital content (soft tissue, fat, muscles) with $m$ for the eyeball mass and $c$ for the discrete viscous damping coefficient.

$$k_{z1} = F_{z1}/w_1 - m(\ddot{w}_1/w_1) - c(\dot{w}_1/w_1) \quad \text{(a)} \qquad k_{z1} \approx F_{z1}/w_1 \quad \text{(b)} \tag{1}$$

In agreement with kinematics governing equations, the peak of maximal displacement $w_1$ was characterized by a nil instantaneous velocity ($\dot{w}_1 \approx 0$) and a non-nil but limited acceleration ($\ddot{w}_1 \neq 0$). The damping term essentially conditioned the time delay between force and displacement signals. The inertia force was very limited since it derived both from the multiplication of the acceleration by the very weak mass $m$ and from the division by the maximal displacement $w_1$.

Finally, the real-time measurement of the peak of maximum displacement $w_1$ and of the simultaneous collinear force $F_{z1}$ provided a first evaluation of the *quasi-static* stiffness $k_{z1}$ of the orbital content as shown in equation (1b).

*2.2 Clinical study*

The protocol was evaluated in a pilot study involving one voluntary patient in accordance with the hospital welfare regulations and guidelines. The patient showed a bilateral proptosis, specifically accentuated in the right eye. A Computed Tomography scan (CT scan) showing the decompressed tissue moving towards the ethmoid sinus region is plotted in Figure 1c.

The measurement process was included into a normal clinical setting and it induced no significant modification of the surgery procedure. The force and displacement measurement process was as follows: two series of data acquisitions were done on the left orbit and one sequence on the right orbit. For each series, data recording was achieved before and after the wall osteotomies and three successive loads were applied by the surgeon.

**3. Results**

A typical data set comprising recorded displacements and forces is shown in Figure 2. It shows the intervention on the patient left eye before the osteotomy (Figure 2a) to quantify the preoperative behavior and after the osteotomy (Figure 2b) to evaluate the tissue decompression. The mean value and standard deviation of displacement $w_1$, force $F_{z1}$ and quasi-static stiffness $k_{z1}$ are given in Table 1. The average value for the whole experiment was 16 N (standard deviation SD: 3 N) for the force and 9 mm (SD: 4 mm) for the eyeball backward displacement. The averaged quasi-static stiffness of the orbital content was evaluated to 2.4 N/mm (SD: 1.2 N/mm).

In term of applied force, the signal was very repetitive in time and magnitude. For each test, the duration of the test sequence was almost 10 s in which three loads of 2 s were applied. The force showed a maximum magnitude up to 20 N and standard deviation lower than 4 N. The mean force in the post-osteotomy phase showed a significant decrease: -35% for the left eye and -11% for the right eye. As shown in Figure 2, repetitive oscillations of signal were depicted during loads exerted by the surgeon (*load 1, 2, 3*) and this observation was made for each eye and both before and after wall osteotomies.

In term of displacement, a mean value up to 11 mm and standard deviation of 5.6 mm were recorded. The displacement signal was noisier than the force one and both signals were globally. It was particularly obvious for the post-osteotomy results. Both for pre-operative and post-operative data, the pumping effect of the successive transient loads induced an increasing backward displacement of the eye-ball.

The quasi-static stiffness was evaluated using the simplified method presented in section 2.1.2. To derive displacements, an interpolation of the signal (polynomial order 4) was made in the peak zone located by grey zones in the typical results of Figure 2. The curve apex provided the maximum displacement $w_1$ as far as the associated time. The first and second polynomial derivatives allowed the velocity and the acceleration to be computed. As previously assumed (section 2.1.2) null velocities were obtained and accelerations were comprised between -10 mm/s$^2$ and -80 mm/s$^2$. The stiffness was derived using equation (1b) using the simultaneous force component $F_{z1}$ and previously verifying that the residual terms of equation (1a) were negligible. Mean value of $k_{z1}$ were comprised between 1.3 N/mm and 4 N/mm showing a standard deviation between 0.7 N/mm and 1.6 N/mm. The damping forces ($\dot{w}_1 \approx 0$) were null and the contribution of inertia forces were negligible. Using the inertia term in equation (1a), the maximal value of 40 x 10$^{-6}$ N/mm was obtained for $\ddot{w}_1 \approx -80 \, \text{mm/s}^2$, $\dot{w}_1 \approx 15 \text{mm}$ and the eyeball mass $m$ of 7g.

## 4. Discussion and conclusion

Combining a mixed approach involving experimental techniques and a clinical pilot study allowed the clinical gesture in proptosis reduction to be investigated. For the first time, the simultaneous measurement of the force and the resulting kinematics allowed data to be quantified intra-operatively. The stiffness of the orbital content was quantified to evaluate the surgery outcome. These results were in agreement with our initial hypothesis.

The force exerted by the surgeon on the eyeball was quantified in a clinical setting with satisfactory accuracy. Very repetitive and not very noisy signals were recorded. The Polaris system allowed the collinear translation to be determined. Measuring 3D real-time displacements in a rather confined area of an operating room was challenging and the displacement signals were much more disturbed. The standard deviation was greater than that of the force because maintaining permanent contact between the device and the corneal protection shell as far as constant orientation was difficult for the surgeon. The ergonomics will be improved in the next version of the device; the use of a non-invasive guide fixed on the patient head has been envisaged.

Oscillations inside each load case (Figure 2) were depicted before and after surgery and for each eye. To find an interpretation of these variations, a default of the force gage was envisaged, but the preliminary validation procedure managed in test bench eliminated this proposal. When pre-operative and post-operative results were compared (figure 2), the synchronicity of displacements referring to the force were more obvious post-operatively when the tissue had been decompressed. So the force oscillations could be explained by a re-arranging of the orbital content (optical nerve, muscle, fluid transfer) particularly sensitive to the pathology level. A third possible explanation was the proprioceptive-motor integration of hand-eye coordination of the operator which could modify his gesture according to feedback through the force gage.

The resulting backward displacements showed a trend to increase during pumping effect due to loading transient application. As shown in Figure 2, the increase was significant after the first loading before surgery and more progressive post-operatively. This trend was found for all measurements (left eye, right eye, two successive loading applications and acquisition sequence). The loads induced an immediate and almost constant rearrangement pre-operatively because of the occlusion of the orbit. On the contrary, the wall opening (Figure 1c) seemed to permit a more monotone evolution of the mechanical behavior of the orbital content essentially due to progressive pressure relaxation and fluid transfer towards the ethmoid sinus region.

This observation was completed by results in term of quasi-static stiffness evaluated during the clinical procedure. The post-treatment of value presented in Table 1, showed that the pre-operative stiffness increase was +48% for the right eye showing an accentuated proptosis. After surgery, this differential value increased up to +60%. The surgery resulted in a stiffness decrease of -48% for the left eye and -43% for the right eye. It was noticed that the poroelastic FE predictive model previously published [Luboz et al., 2004, 2005] was in good agreement with the clinical data of the present study since comprised in the range of standard variation of the experimental results.

Despite these encouraging results, it was considered that the stiffness evaluation in quasi-static domain around the amplitude peaks was a first approximation. The interpolation of displacement around displacement peaks, successive velocity and acceleration induced noticeable discrepancies which enlarge standard deviation. Some reasonable theoretical assumptions were made concerning the limited influence of inertia and damping terms, but the dissipated energy seemed to play a role on the phase angle between force and displacement. To answer this, the next step of the study will be to use the overall displacement and force signals as input measures into a non-linear inverse algorithm to identify the equivalent rheological model of the orbital content.

Finally, the clinical pilot study proved that the quantification of the surgical gesture was realistic in a clinical setting and this result was particularly clinical relevant in orbital surgery. The original device and associated protocol also showed a significant interest to help in the diagnosis of the pathology level and also in the treatment follow-up before surgery. The measured force and displacement to update predictive numerical models could be very pertinent for surgery planning and intra-operative guidance to conduct minimally invasive surgery. The precondition to all these promising projects will be to enlarge the patient database.

**References**

Adenis J. P., Robert P. Y. 1994, Décompression orbitaire selon la technique d'Olivari. J. Fr. Ophtalmol. 17, 686-691.

Gerard J.M., Ohayon J., Luboz V., Perrier P. & Payan Y., 2005. Non linear elastic properties of the lingual and facial tissues assessed by indentation technique. Application to the biomechanics of speech production. Medical Engineering & Physics. 27, 884-892.

Gosling T, Westphal R, Faulstich J, Sommer K, Wahl F, Krettek C, Hufner T., 2006. Forces and torques during fracture reduction: Intraoperative measurements in the femur. J Orthop Res. 24, 333-8.

Hendriks F.M., Brokken D., Oomens C.W.J., Bader D.L., Baaijens F.P.T., 2006. The relative contributions of different skin layers to the mechanical behavior of human skin in vivo using suction experiments. Medical Engineering & Physics. 28, 259-266.

Jenkinson RJ, Sanders DW, Macleod MD, Domonkos A, Lydestadt J., 2005. Intraoperative diagnosis of syndesmosis injuries in external rotation ankle fractures. J Orthop Trauma. 19: 604-9.

Luboz V., Pedrono A., Amblard D., Swider P., Payan Y. & Boutault F., 2004. Prediction of tissue decompression in orbital surgery. Clinical Biomechanics. 19, 202-208.

Luboz V., Ambard D., Swider P., Boutault F. & Payan Y., 2005. Computer assisted planning and orbital surgery: patient-related prediction of osteotomy size in proptosis reduction. Clinical Biomechanics. 20, 900-905.

Miller K, Chinzei K., Orssengo G., Bednarz P., 2000. Mechanical properties of brain tissue in-vivo: experiment and computer simulation. Journal of Biomechanics. 33, 1369-1376.

Ottensmeyer M., Kerdok A., Howe R., Dawson S., 2004. The Effects of Testing Environment on the Viscoelastic Properties of Soft Tissues. Lecture Notes in Computer Science. 3078, 9-18.

Payan Y., editor, 2005. Biomechanics Applied to Computer Assisted Surgery. Research Signpost Publisher, ISBN 81-308-0031-4.

Payan Y., Luboz V., Swider P., Amblard D., 2003. Outil et procédé de mesure de la raideur mécanique d'un milieu selon une direction déterminée. Université Joseph Fourier de Grenoble / Université Paul Sabatier de Toulouse. Patent 03/51109, France.

Saraux H., Biais B., Rossazza C., 1987. Ophtalmologie. Masson editor, 341-353.

Stanley R.J., McCaffrey T.V., Offord K.P., DeSanto L.W., 1989. Superior and transantral orbital decompression procedures. Effects on increased intraorbital pressure and orbital dynamics. Arch. Otolaryngol. Head Neck Surg. 115, 369-373.

Taylor R., Lavallée S., Burdea G. & Mosges R., 1996. Computer integrated surgery: Technology and clinical applications. Cambridge, MA: MIT Press.

Viceconti M, Olsen S, Nolte LP, Burton K., 2005. Extracting clinically relevant data from finite element simulations. Clin Biomech. 20: 451-4.

*Intra-operative quantification of the surgical gesture in orbital surgery: application to the proptosis reduction. V. Luboz, P. Swider, D. Ambard, F. Boutault, Y. Payan*

**Captions for table and illustrations**

**Table 1** - Intra-operative results of load $F_{z1}$(N), eyeball backward displacement $w_1$(mm), and mean quasi-static stiffness $K_{z1}$(N/mm) of the orbital content of the patient.

**Figure 1** - Clinical procedure and experimental device: (a) intra-operative positioning; (b) technological scheme: (1) aluminum conic base, (2) stainless membrane involving strain gages, (3) stainless ring, (4) stainless shell, (5) silicon cylinder, (6) polymer rigid body for navigation, (7) wires; (c) CT scan of the proptosis and image segmentation

**Figure 2** - Intra-operative quantification of the clinical gesture for the left orbit of the patient: (a) force and displacement before surgery; (b) force and displacement after surgery
(—): force $F_{z1}$, (- - -): displacement $w_1$

*Intra-operative quantification of the surgical gesture in orbital surgery: application to the proptosis reduction. V. Luboz, P. Swider, D. Ambard, F. Boutault, Y. Payan*

|  | Pre osteotomy | | | Post osteotomy | | |
| --- | --- | --- | --- | --- | --- | --- |
|  | $F_{z1}(N)$ | $w_1(mm)$ | $k_{z1}(N/mm)$ | $F_{z1}(N)$ | $w_1(mm)$ | $k_{z1}(N/mm)$ |
| left orbit | 17.4 *SD 3.9* | 8.7 *SD 3.8* | 2.5 *SD 1.6* | 11.2 *SD 3.5* | 9.4 *SD 2.7* | 1.3 *SD 0.7* |
| right orbit | 18.2 *SD 2.2* | 6.7 *SD 3.9* | 3.7 *SD 1.4* | 16.2 *SD 1.7* | 10.2 *SD 5.6* | 2.1 *SD 1.3* |

**Table 1 -** Intra-operative results of load $F_{z1}$(N), eyeball backward displacement $w_1$(mm), and mean quasi-static stiffness $K_{z1}$(N/mm) of the orbital content of the patient.

*Intra-operative quantification of the surgical gesture in orbital surgery: application to the proptosis reduction.    V. Luboz, P. Swider, D. Ambard, F. Boutault, Y. Payan*

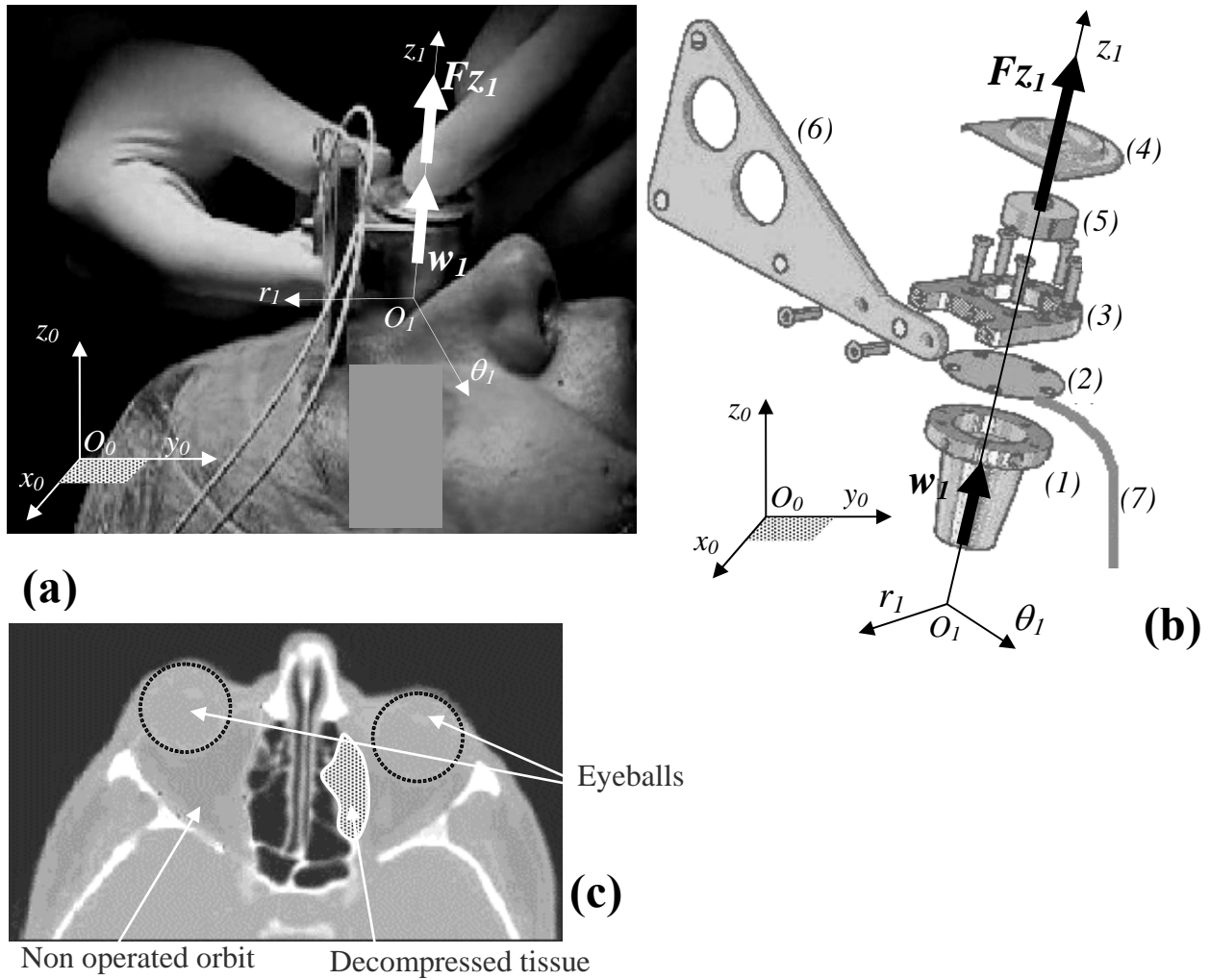

**Figure 1 -** Clinical procedure and experimental device: (a) intra-operative positioning; (b) technological scheme: (1) aluminum conic base,  (2) stainless steel membrane involving strain gages, (3) stainless ring, (4) stainless steel shell, (5) silicon cylinder, (6) polymer rigid body for navigation, (7) wires; (c) CT scan of the proptosis and image segmentation

*Intra-operative quantification of the surgical gesture in orbital surgery: application to the proptosis reduction. V. Luboz, P.Swider, D. Ambard, F. Boutault, Y. Payan*

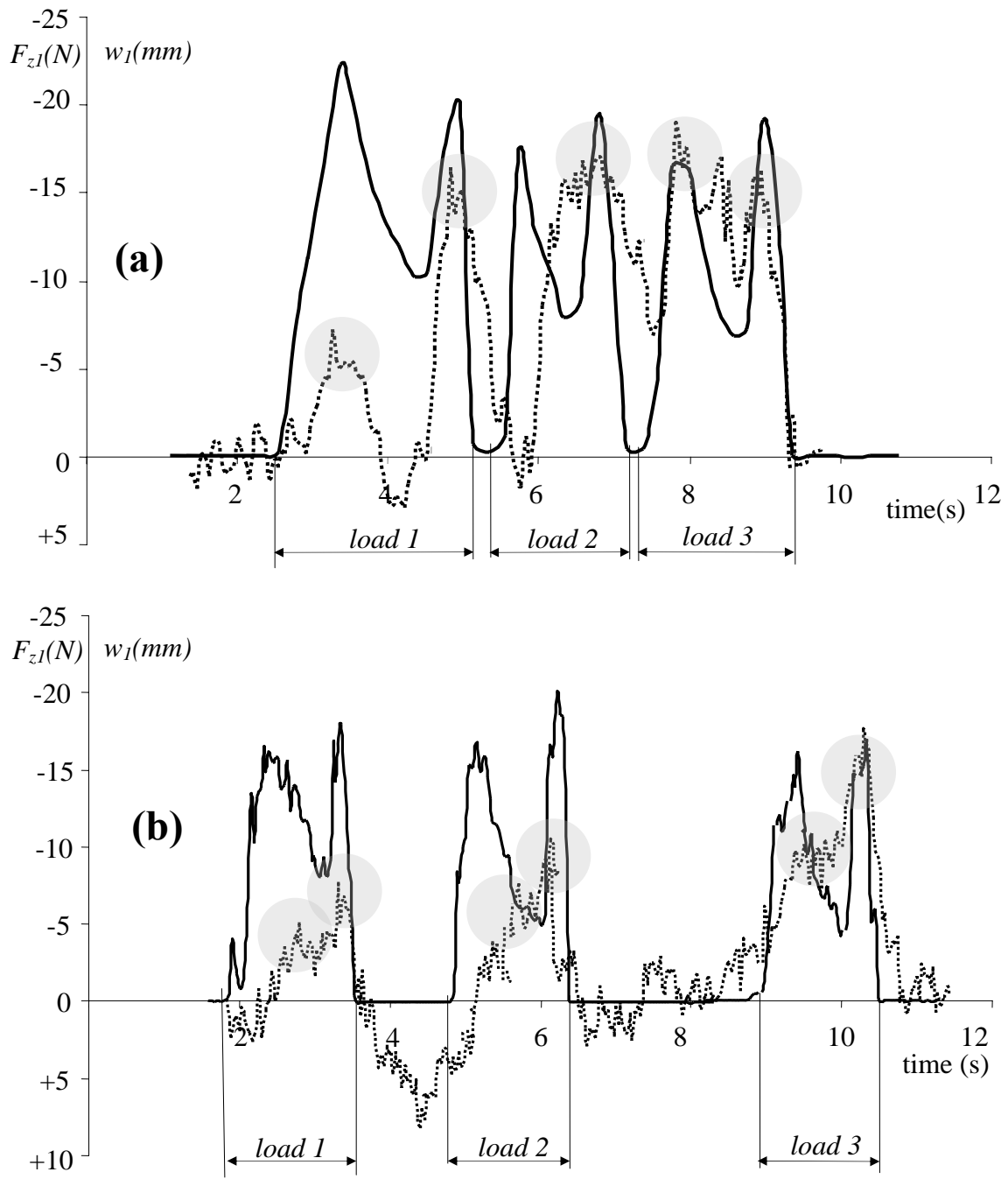

**Figure 2 -** Intra-operative quantification of the clinical gesture for the left orbit of the patient: (a) force and displacement before surgery; (b) force and displacement after surgery
(—): force $F_{z1}$, (- - -): displacement $w_1$